\documentstyle[aps,epsf]{revtex}

\begin{document}
\title{
Viscous Bianchi type I Universes in Brane Cosmology}
\author{T. Harko\thanks{%
E-mail: tcharko@hkusua.hku.hk} and M. K. Mak\thanks{E-mail: mkmak@vtc.edu.hk}}
\address{Department of Physics, The University of Hong Kong, Pokfulam Road, Hong Kong}
\maketitle

\begin{abstract}
We consider the dynamics of a viscous cosmological fluid in the generalized
Randall-Sundrum model for an anisotropic, Bianchi type I brane. To describe
the dissipative effects we use the Israel-Hiscock-Stewart full causal
thermodynamic theory. By assuming that the matter on the brane obeys a
linear barotropic equation of state, and the bulk viscous pressure has a
power-law dependence on the energy density, the general solution of the
field equations can be obtained in an exact parametric form. The obtained
solutions describe generally a non-inflationary brane world. In the large
time limit the brane Universe isotropizes, ending in an isotropic and
homogeneous state. The evolution of the temperature and of the comoving
entropy of the Universe is also considered, and it is shown that due to
the viscous dissipative processes a large amount of entropy is created in the
early stages of evolution of the brane world.
\end{abstract}


\section{Introduction}

The idea \cite{RS99a,RS99b} that our four-dimensional Universe might be a
three-brane embedded in a higher dimensional space-time has attracted much
attention. According to the brane-world scenario, the physical fields in our
four-dimensional space-time, which are assumed to arise as fluctuations of
branes in string theories, are confined to the three brane, while gravity
can freely propagate in the bulk space-time, with the gravitational
self-couplings not significantly modified. The model originated from the
study of a single $3$-brane embedded in five dimensions, with the $5D$
metric given by $ds^{2}=e^{-f(y)}\eta _{\mu \nu }dx^{\mu }dx^{\nu }+dy^{2}$,
which can produce a large hierarchy between the scale of particle physics
and gravity due to the appearance of the warp factor. Even if the fifth
dimension is uncompactified, standard $4D$ gravity is reproduced on the
brane. Hence this model allows the presence of large or even infinite
non-compact extra dimensions and our brane is identified to a domain wall in
a $5$-dimensional anti-de Sitter space-time.

The Randall-Sundrum model was inspired by superstring theory. The
ten-dimensional $E_{8}\times E_{8}$ heterotic string theory, which contains
the standard model of elementary particle, could be a promising candidate
for the description of the real Universe. This theory is connected with an
eleven-dimensional theory, M-theory, compactified on the orbifold $%
R^{10}\times S^{1}/Z_{2}$ \cite{HW96}. In this model we have two separated
ten-dimensional manifolds. For a review of dynamics and geometry of brane
Universes see \cite{Ma01}.

The static Randall-Sundrum solution has been extended to time-dependent
solutions and their cosmological properties have been extensively studied 
\cite{KK00,BDEL00,BDL00,STW00,LMW00,KIS00,BDBL00,CEHS00,MSS00,To01,F01}. In
one of the first cosmological applications of this scenario, it was pointed
out that a model with a non-compact fifth dimension is potentially viable,
while the scenario which might solve the hierarchy problem predicts a
contracting Universe, leading to a variety of cosmological problems \cite
{CsGrKoTe99}. By adding cosmological constants to the brane and bulk, the
problem of the correct behavior of the Hubble parameter on the brane has
been solved by Cline, Grojean and Servant \cite{ClGrSe99}. As a result one
also obtains normal expansion during nucleosynthesis, but faster than normal
expansion in the very early Universe. The creation of a spherically
symmetric brane-world in AdS bulk has been considered, from a quantum
cosmological point of view, with the use of the Wheeler-de Witt equation, by
Anchordoqui, Nunez and Olsen \cite{AnNuOl00}.

The effective gravitational field equations on the brane world, in which all
the matter forces except gravity are confined on the $3$-brane in a $5$%
-dimensional space-time with $Z_{2}$-symmetry have been obtained, by using a
geometric approach, by Shiromizu, Maeda and Sasaki \cite{SMS00,SSM00}. The
correct signature for gravity is provided by the brane with positive
tension. If the bulk space-time is exactly anti-de Sitter, generically the
matter on the brane is required to be spatially homogeneous. The electric
part of the $5$-dimensional Weyl tensor $E_{IJ}$ gives the leading order
corrections to the conventional Einstein equations on the brane. The
four-dimensional field equations for the induced metric and scalar field on
the world-volume of a $3$-brane in the five-dimensional bulk with Einstein
gravity plus a self-interacting scalar field have been derived by Maeda and
Wands \cite{MW00}.

The linearized perturbation equations in the generalized Randall-Sundrum
model have been obtained, by using the covariant nonlinear dynamical
equations for the gravitational and matter fields on the brane, by Maartens 
\cite{Ma00}. A systematic analysis, using dynamical systems techniques, of
the qualitative behavior of the Friedmann-Robertson-Walker (FRW), Bianchi
type I and V cosmological models in the Randall-Sundrum brane world
scenario, with matter on the brane obeying a barotropic equation of state $%
p=(\gamma -1)\rho $, has been realized by Campos and Sopuerta \cite
{CS01,CS201}. In particular, they constructed the state spaces for these
models and discussed what new critical points appear, the occurrence of
bifurcations and the dynamics of the anisotropy for both a vanishing and
non-vanishing Weyl tensor in the bulk.

The general exact solution of the field equations for an anisotropic brane
with Bianchi type I and V geometry, with perfect fluid and scalar fields as
matter sources has been found in \cite{ChHaMa01a}. Expanding Bianchi type I
and V brane-worlds always isotropize, although there could be intermediate
stages in which the anisotropy grows. In spatially homogeneous brane world
cosmological models the initial singularity is isotropic and hence the
initial conditions problem is solved \cite{Co01a}. Consequently, these
models do not exhibit Mixmaster or chaotic-like behavior close to the
initial singularity \cite{Co01b}. Other properties of brane world
cosmologies have been considered in \cite{F01} and \cite{SaVeFe01}.

Realistic brane-world cosmological models require the consideration of more
general matter sources to describe the evolution and dynamics of the very
early Universe. Limits on the initial anisotropy induced by the
5-dimensional Kaluza-Klein graviton stresses by using the CMB anisotropies
have been obtained by Barrow and Maartens \cite{BaMa01} and Leong, Challinor, 
Maartens and Lasenby \cite{LeChMaLa02}. Anisotropic Bianchi type I
brane-worlds with a pure magnetic field and a perfect fluid have also been
analyzed \cite{BaHe01}. Rotational perturbations of brane world cosmological
models have been studied in \cite{ChHaKaMa02}.

Most of the investigations of brane cosmological models have been performed
under the simplifying assumption of a perfect cosmological fluid. But in
many cosmological situations an idealized fluid model of matter is
inappropriate, especially in the case of matter at very high densities and
pressures. Such possible situations are the relativistic transport of
photons, mixtures of cosmic elementary particles, evolution of cosmic
strings due to their interaction with each other and surrounding matter,
classical description of the (quantum) particle production phase,
interaction between matter and radiation, quark and gluon plasma viscosity
etc. From a physical point of view the inclusion of dissipative terms in the
energy-momentum tensor of the cosmological fluid seems to be the best
motivated generalization of the matter term of the gravitational field
equations.

The first attempts at creating a theory of relativistic dissipative fluids
were those of Eckart \cite{Ec40} and Landau and Lifshitz \cite{LaLi87}.
These theories are now known to be pathological in several respects.
Regardless of the choice of equation of state, all equilibrium states in
these theories are unstable and in addition signals may be propagated
through the fluid at velocities exceeding the speed of light. These problems
arise due to the first order nature of the theory, that is, it considers
only first-order deviations from the equilibrium leading to parabolic
differential equations, hence to infinite speeds of propagation for heat
flow and viscosity, in contradiction with the principle of causality.
Conventional theory is thus applicable only to phenomena which are
quasi-stationary, i.e. slowly varying on space and time scales characterized
by mean free path and mean collision time.

A relativistic second-order theory was found by Israel \cite{Is76} and
developed by Israel and Stewart \cite{IsSt76} into what is called
``transient'' or ``extended'' irreversible thermodynamics. In this model
deviations from equilibrium (bulk stress, heat flow and shear stress) are
treated as independent dynamical variables, leading to a total of 14
dynamical fluid variables to be determined. However, Hiscock and Lindblom 
\cite{HiLi89} and Hiscock and Salmonson \cite{HiSa91} have shown that most
versions of the causal second order theories omit certain divergence terms.
The truncated causal thermodynamics of bulk viscosity leads to pathological
behavior in the late Universe, while the solutions of the full causal theory
are \cite{AnPaRo98}: a) for stable fluid configurations the dissipative
signals propagate causally, b) unlike in Eckart-type's theories, there is no
generic short wave-length secular instability and c) even for rotating
fluids, the perturbations have a well-posed initial value problem. For
general reviews on causal thermodynamics and its role in relativity see \cite
{Ma96} and\cite{Ma95}.

Causal bulk viscous thermodynamics has been extensively used for describing
the dynamics and evolution of the early Universe or in an astrophysical
context. But due to the complicated character of the evolution equations,
very few exact cosmological solutions of the gravitational field equations
are known in the framework of the full causal theory. For a homogeneous
Universe filled with a full causal viscous fluid source obeying the relation 
$\xi \sim \rho ^{1/2}$, exact general solutions of the field equations have
been obtained in \cite{ChJa97}, \cite{MaHa98}, \cite{MaHa98a}, \cite{MaHa99}%
. In this case the evolution of the bulk viscous cosmological model can be
reduced to a Painleve-Ince type differential equation, whose invariant form
can be reduced, by means of non-local transformations, to a linear
inhomogeneous ordinary second-order differential equation with constant
coefficients \cite{Ch97}. It has also been proposed that causal bulk viscous
thermodynamics can model on a phenomenological level matter creation in the
early Universe \cite{MaHa98}, \cite{MaHa99a}. Exact causal viscous
cosmologies with $\xi \sim \rho ^{s}$ have been obtained in \cite{HaMa99}.

Because of technical reasons, most investigations of dissipative causal
cosmologies have assumed FRW symmetry (i.e. homogeneity and isotropy) or
small perturbations around it \cite{MaTr97}. The Einstein field equations
for homogeneous models with dissipative fluids can be decoupled and
therefore are reduced to an autonomous system of first order ordinary
differential equations, which can be analyzed qualitatively \cite{CoHo95}, 
\cite{CoHo96}.

The influence of the bulk viscosity of the matter on the brane has been
analyzed, for an isotropic flat FRW geometry, in \cite{ChHaMa01b}.
Dissipative viscous effects in the brane world lead to important differences
in the cosmological dynamics, as compared to the standard general
relativistic cosmology. Since the effects of the extra-dimensions and also
the viscous effects are more important at high matter densities, the most
important contribution to the energy density of the matter in this regime
comes from the quadratic term in density. Consequently, during the early
period of evolution of the brane world the energy density of matter is
proportional to the Hubble parameter, in opposition to the standard general
relativistic case with energy density proportional to the square of the
Hubble parameter.

It is the purpose of the present paper to investigate the effects of the
bulk viscosity of the cosmological matter fluid on the dynamics of an
anisotropic, Bianchi type I brane world. In order to solve the field
equations, we assume a specific equation of state for the bulk viscous
pressure. With this choice the general solution of the field equations can
be expressed in an exact parametric form. The obtained solution describes a
non-inflationary Universe, tending in a large time limit to an isotropic
homogeneous geometry. The behavior of the temperature and comoving entropy
is also considered.

The present paper is organized as follows. The field equations on the brane
describing the evolution of a viscous cosmological fluid are written down in
Section II. In Section III we present the general solution of the field
equations in the case of a power-law dependence of the bulk viscous pressure
on the energy density. In Section IV we discuss and conclude our results.

\section{Dissipative cosmological fluids on the anisotropic brane}

In the $5D$ space-time the brane-world is located as $Y(X^{I})=0$, where $%
X^{I},\,I=0,1,2,3,4$ are $5$-dimensional coordinates. The effective action
in five dimensions is \cite{MW00} 
\begin{equation}
S=\int d^{5}X\sqrt{-g_{5}}\left( \frac{1}{2k_{5}^{2}}R_{5}-\Lambda
_{5}\right) +\int_{Y=0}d^{4}x\sqrt{-g}\left( \frac{1}{k_{5}^{2}}K^{\pm
}-\lambda +L^{\text{matter}}\right) ,
\end{equation}
with $k_{5}^{2}=8\pi G_{5}$ the $5$-dimensional gravitational coupling
constant and where $x^{\mu },\,\mu =0,1,2,3$ are the induced $4$-dimensional
brane world coordinates. $R_{5}$ is the $5D$ intrinsic curvature in the bulk
and $K^{\pm }$ is the intrinsic curvature on either side of the brane.

On the $5$-dimensional space-time (the bulk), with the negative vacuum
energy $\Lambda_5$ and brane energy-momentum as source of the gravitational
field, the Einstein field equations are given by 
\begin{equation}
G_{IJ} = k_5^2 T_{IJ}, \qquad T_{IJ} = -\Lambda_5 g_{IJ} + \delta(Y) \left[
-\lambda g_{IJ} + T_{IJ}^{\text{matter}} \right].
\end{equation}

In this space-time a brane is a fixed point of the $Z_2$ symmetry. In the
following capital Latin indices run in the range $0,...,4$ while Greek
indices take the values $0,...,3$.

Assuming a metric of the form $ds^{2}=(n_{I}n_{J}+g_{IJ})dx^{I}dx^{J}$, with 
$n_{I}dx^{I}=d\chi $ the unit normal to the $\chi =\text{const.}$
hypersurfaces and $g_{IJ}$ the induced metric on $\chi =\text{const.}$
hypersurfaces, the effective four-dimensional gravitational equations on the
brane (which are the consequence of the Gauss-Codazzi equations) take the
form \cite{SMS00,SSM00}: 
\begin{equation}
G_{\mu \nu }=-\Lambda g_{\mu \nu }+k_{4}^{2}T_{\mu \nu }+k_{5}^{4}S_{\mu \nu
}-E_{\mu \nu },  \label{Ein}
\end{equation}
where $S_{\mu \nu }$ is the local quadratic energy-momentum correction 
\begin{equation}
S_{\mu \nu }=\frac{1}{12}TT_{\mu \nu }-\frac{1}{4}T_{\mu }{}^{\alpha }T_{\nu
\alpha }+\frac{1}{24}g_{\mu \nu }\left( 3T^{\alpha \beta }T_{\alpha \beta
}-T^{2}\right) ,
\end{equation}
and $E_{\mu \nu }$ is the non-local effect from the bulk free gravitational
field, transmitted via the projection of the bulk Weyl tensor $C_{IAJB}$: $%
E_{IJ}=C_{IAJB}n^{A}n^{B}$ and $\ E_{IJ}\rightarrow E_{\mu \nu }\delta
_{I}^{\mu }\delta _{J}^{\nu }$ as $\chi \rightarrow 0$. The four-dimensional
cosmological constant, $\Lambda $, and the gravitational coupling constant
on the brane, $k_{4}$, are given by $\Lambda =k_{5}^{2}\left( \Lambda _{5}+%
\frac{k_{5}^{2}\lambda ^{2}}{6}\right) /2$ and $k_{4}^{2}=k_{5}^{4}\lambda
/6 $, respectively.

The Einstein equation in the bulk, the Codazzi equation, also implies the
conservation of the energy momentum tensor of the matter on the brane, 
\begin{equation}
D_{\nu }T_{\mu }{}^{\nu }=0.  \label{DT}
\end{equation}
Moreover, the contracted Bianchi identities imply that the projected Weyl
tensor should obey the constraint 
\begin{equation}
D_{\nu }E_{\mu }{}^{\nu }=k_{5}^{4}D_{\nu }S_{\mu }{}^{\nu }.  \label{DE}
\end{equation}
Finally, the equations (\ref{Ein})-(\ref{DT}) and (\ref{DE}) give the
complete set of field equations for the brane gravitational field.

For any matter fields (scalar field, perfect fluids, kinetic gases,
dissipative fluids etc.) the general form of the brane energy-momentum
tensor can be covariantly given as \cite{Ma96} 
\begin{equation}
T_{\mu \nu }=\rho u_{\mu }u_{\nu }+p_{eff}h_{\mu \nu }+\pi _{\mu \nu
}+2q_{(\mu }u_{\nu )}.  \label{EMT}
\end{equation}
The decomposition is irreducible for any chosen $4$-velocity $u^{\mu }$.
Here $\rho $ and $p_{eff}$ are the energy density and the effective
isotropic pressure, and $h_{\mu \nu }=g_{\mu \nu }+u_{\mu }u_{\nu }$
projects orthogonal to $u^{\mu }$. The energy flux $q_{\mu }$ and the
anisotropic stress $\pi _{\mu \nu }$ obey the conditions $q_{\mu }=q_{<\mu
>} $ and $\pi _{\mu \nu }=\pi _{<\mu \nu >}$, respectively, where angular
brackets denote the projected, symmetric and trace-free parts: 
\begin{equation}
V_{<\mu >}=h_{\mu }{}^{\nu }V_{\nu },\qquad W_{<\mu \nu >}=\left[ h_{(\mu
}{}^{\alpha }h_{\nu )}{}^{\beta }-\frac{1}{3}h^{\alpha \beta }h_{\mu \nu }%
\right] W_{\alpha \beta }.
\end{equation}

We assume that the equilibrium thermodynamics pressure $p$ of the
cosmological fluid obeys a linear barotropic equation of state $p=\left(
\gamma -1\right) \rho $, $\gamma \in \left[ 1,2\right] $ and $\gamma =$%
constant.

The effect of the bulk viscosity of the cosmological fluid on the brane can
be considered by adding to the usual thermodynamic pressure $p$ the bulk
viscous pressure $\Pi $ and formally substituting the pressure terms in the
energy-momentum tensor by $p_{eff}=p+\Pi $.

The anisotropic stress $\pi _{\mu \nu }$ of the matter on the brane
satisfies the evolution equation \cite{Ma96}

\begin{equation}
\tau _{2}h_{\alpha }^{\beta }h_{\beta }^{\nu }\dot{\pi}_{\mu \nu }+\pi
_{\alpha \beta }=-2\eta \sigma _{\alpha \beta }-\left[ \eta T\left( \frac{%
\tau _{2}}{2\eta T}u^{\nu }\right) _{;\nu }\pi _{\alpha \beta }\right] ,
\end{equation}
where $\eta $ is the shear viscosity coefficient, $\tau _{2}=2\eta \beta
_{2} $, with $\beta _{2}$ the thermodynamic coefficient for the tensor
dissipative contribution to the entropy density and $\sigma _{\alpha \beta }$
is the shear tensor. In the following we assume that the main contribution
to entropy generation is via scalar dissipation, that is, we consider that
the dissipative contribution from the shear viscosity can be neglected, $%
\eta \approx 0$. Consequently, the anisotropic stresses of the matter on the
brane also vanish, $\pi _{\mu \nu }\approx 0$. We consider that in Eq. (\ref
{EMT}) the heat transfer is zero, that is, we take $q_{\mu }=0$.

Then the quadratic corrections to the matter energy momentum tensor on the
brane are given by \cite{SMS00} 
\begin{equation}
S_{\mu \nu }=\frac{1}{12}\rho ^{2}u_{\mu }u_{\nu }+\frac{1}{12}\rho (\rho
+2p_{eff})h_{\mu \nu }.
\end{equation}

The symmetry properties of $E_{\mu \nu }$ imply that in general we can
decompose it irreducibly with respect to a chosen $4$-velocity field $u^{\mu
}$ as 
\begin{equation}
E_{\mu \nu }=-k^{4}\left[ {\cal U}\left( u_{\mu }u_{\nu }+\frac{1}{3}h_{\mu
\nu }\right) +{\cal P}_{\mu \nu }+2{\cal Q}_{(\mu }u_{\nu )}\right] ,
\label{WT}
\end{equation}
where $k=k_{5}/k_{4}$. The non-local anisotropic fields ${\cal P}_{\mu \nu }$
and ${\cal Q}_{\mu }$ include five-dimensional gravitational wave modes. To
determine the brane dynamics one must solve the 5D field equations
completely, by choosing appropriate boundary conditions in the bulk. For the
only known solution of the field equations in the bulk, the Schwarzschild-
anti de Sitter bulk that contains a moving brane, the contribution of these
terms is zero \cite{BDEL00}. The effects on the CMB radiation anisotropy of
the non-local stresses under various plausible physical assumptions have
been discussed in \cite{BaMa01} and \cite{LeChMaLa02}. One such possibility
is to assume that the time evolution of the non-local stress is proportional
to the energy density of the anisotropic source $\rho ^{kk}$, with $\rho
^{kk}$ perturbatively small relative to the matter energy density $\rho $, $%
\rho ^{kk}<<\rho $. Since, on the other hand, the viscous dissipative
effects are proportional to the matter density, their effects will generally dominate
the non-local effects from the bulk. Therefore we shall also assume a
vanishing effective non-local energy density, ${\cal P}_{\mu \nu }\approx
0\approx {\cal Q}_{\mu }$.

The line element of a Bianchi type I space-time, which generalizes the flat
Friedmann-Robertson-Walker metric to the anisotropic case, is given by 
\begin{equation}
ds^{2}=-dt^{2}+a_{1}^{2}\left( t\right) dx^{2}+a_{2}^{2}\left( t\right)
dy^{2}+a_{3}^{2}\left( t\right) dz^{2}.  \label{5}
\end{equation}

We define the following variables: 
\begin{equation}
V=\prod_{i=1}^{3}a_{i},3H=\sum_{i=1}^{3}H_{i},H_{i}=\frac{\dot{a}_{i}}{a_{i}}%
,\Delta H_{i}=H_{i}-H,i=1,2,3.  \label{6}
\end{equation}

In Eqs. (\ref{6}) $V$ is the volume scale factor, $H_{i},i=1,2,3$ are the
directional Hubble parameters, and $H$ is the mean Hubble parameter. From
Eqs. (\ref{6}) we also obtain $H=\dot{V}/3V$.

The physical quantities of observational interest in cosmology are the
expansion scalar $\theta $, the mean anisotropy parameter $A$ and the shear
scalar $\sigma ^{2}$, which are defined according to 
\begin{equation}
\theta =3H,3A=\sum_{i=1}^{3}\left( \frac{\Delta H_{i}}{H}\right)
^{2},2\sigma ^{2}=\sigma _{ik}\sigma
^{ik}=\sum_{i=1}^{3}H_{i}^{2}-3H^{2}=3AH^{2}.  \label{10}
\end{equation}

Using the variables (\ref{6}), the Einstein gravitational field equations,
the Bianchi identity and the evolution equation for the non-local dark
energy take the form: 
\begin{eqnarray}
3\dot{H}+\sum_{i=1}^{3}H_{i}^{2} &=&\Lambda +\frac{k_{4}^{2}}{2}\left[
\left( 2-3\gamma \right) \rho -3\Pi \right] +\frac{k_{4}^{2}\rho }{2\lambda }%
\left[ \left( 1-3\gamma \right) \rho -3\Pi \right] -\frac{6{\cal U}}{%
k_{4}^{2}\lambda },  \label{dH} \\
\frac{1}{V}\frac{d}{dt}\left( VH_{i}\right) &=&\Lambda +\frac{k_{4}^{2}}{2}%
\left[ \left( 2-\gamma \right) \rho -\Pi \right] +\frac{k_{4}^{2}\rho }{%
2\lambda }\left[ \left( 1-\gamma \right) \rho -\Pi \right] +\frac{2{\cal U}}{%
k_{4}^{2}\lambda },i=1,2,3,  \label{dVHi} \\
\dot{\rho}+3\gamma H\rho &=&-3H\Pi ,  \label{drho} \\
\dot{{\cal U}}+4H{\cal U} &=&0.  \label{dU}
\end{eqnarray}

The causal evolution equation for the bulk viscous pressure $\Pi $ is given
by \cite{Ma95} 
\begin{equation}
\tau \dot{\Pi}+\Pi =-3\xi H-\frac{1}{2}\tau \Pi \left( 3H+\frac{\dot{\tau}}{%
\tau }-\frac{\dot{\xi}}{\xi }-\frac{\dot{T}}{T}\right) ,  \label{bulk}
\end{equation}
where $\tau $ is the relaxation time, $T$ is the temperature and $\xi $ is
the bulk viscosity coefficient.

Eq. (\ref{bulk}) arises as the simplest way (linear in $\Pi $) to satisfy
the $H$-theorem (i.e., for the entropy production to be non-negative, $%
S^{\mu }{}_{;\mu }=\Pi ^{2}/(\xi T)\geq 0$). The particle flow vector $%
N^{\mu }$ is given by $N^{\mu }=nu^{\mu }$, where $n\geq 0$ is the particle
number density. In the framework of causal thermodynamics and limiting
ourselves to second-order deviations from equilibrium, the entropy flow
vector $S^{\mu }$ takes the form $S^{\mu }=eN^{\mu }-\tau \Pi ^{2}u^{\mu
}/2\xi T$, where $e$ is the entropy per particle.

An important observational quantity is the deceleration parameter 
\begin{equation}
q=\frac{d}{dt}\left( \frac{1}{H}\right) -1=\frac{\rho +3p+3\Pi +2\frac{K^{2}%
}{V^{2}}}{2\left( \rho +\frac{K^{2}}{2V^{2}}\right) }.
\end{equation}

The sign of the deceleration parameter indicates whether the model inflates
or not. The positive sign of $q$ corresponds to ``standard'' decelerating
models, whereas the negative sign indicates inflation.

\section{General Solution of the Field Equations}

Integrating (\ref{dU}) yields for the dark matter term the expression ${\cal %
U=U}_{0}V^{-\frac{4}{3}}$, where ${\cal U}_{0}\geq 0$ is an arbitrary
constant of integration.

By adding Eqs.(\ref{dVHi}) we find 
\begin{equation}
\frac{1}{V}\frac{d}{dt}\left( VH\right)=\dot{H}+3H^{2}=\Lambda +\frac{%
k_{4}^{2}}{2}\left[ \left( 2-\gamma \right) \rho -\Pi \right] +\frac{%
k_{4}^{2}\rho }{2\lambda }\left[ \left( 1-\gamma \right) \rho -\Pi \right] +%
\frac{2{\cal U}}{k_{4}^{2}\lambda }.  \label{19}
\end{equation}

Substracting Eq. (\ref{19}) from Eqs. (\ref{dVHi}) we obtain 
\begin{equation}
H_{i}=H+\frac{K_{i}}{V},i=1,2,3,  \label{20}
\end{equation}
with $K_{i},i=1,2,3$ constants of integration satisfying the consistency
condition $\sum_{i=1}^{3}K_{i}=0$. By using Eq. (\ref{20}), it is easy to
show that 
\begin{equation}
\sum_{i=1}^{3}H_{i}^{2}=3H^{2}+\frac{K^{2}}{V^{2}},  \label{21}
\end{equation}
where we denoted $K^{2}=\sum_{i=1}^{3}K_{i}^{2}$.

Usually, in order to find the general solution of the gravitational field
equations, the equations of state of the bulk viscosity coefficient,
temperature and relaxation time are given, in terms of the energy density of
the cosmological fluid. Then the thermodynamic and geometric parameters are
obtained by solving the field equations. But in the present paper we shall
use a different approach, namely we shall assume that the functional
dependence of the bulk viscous pressure on the energy density is known.
Therefore the behavior of the bulk viscosity coefficient, temperature and
relaxation time can be obtained from the field equations. Hence for the bulk
viscous pressure we assume the following general power-law type density
dependence: 
\begin{equation}
\Pi =-\gamma \Pi _{0}\rho ^{m},m=\text{constant}>0,m\neq 1,\Pi _{0}=\text{%
constant}\geq 0.  \label{Pi}
\end{equation}

This form of the bulk viscous pressure is sugested by the form of $\Pi $ in
the non-causal theory, where $\Pi \sim \xi H$. This relation follows by
taking $\tau =0$ in Eq.(\ref{bulk}). Assuming a power law density dependence
of the bulk viscosity coefficient on the density, $\xi \sim \rho ^{s}$, $s=$%
constant$>0$, and taking into account that generally the Hubble parameter
can also be approximated by some power of the energy density, we obtain Eq. (%
\ref{Pi}) in a natural way. By using such a form for the viscous stress
Barrow \cite{Ba87} has obtained a class of homogeneous and isotropic cosmological models which begin
in a de Sitter state but subsequently deflate towards the flat Friedman
model. Barrow \cite{Ba88} has also considered the effects of the anisotropies on the early cosmological evolution,
within the framework of the non-causal thermodynamical approch, for two distinct
classes of models corresponding to the choices $m>1$ and $m<1$ in the expression
of the bulk viscous pressure. In both cases the initial anisotropies are smoothed out
by the inflationary expansion of the Universe.
The range of values of $m$ can be roughly estimated by considering
the isotropic limit of the model. Since for a high density cosmological
fluid $s\leq 1/2$ \cite{BeNiKh79}, and for standard cosmological models $%
H\sim \rho ^{1/2}$, it follows that in conventional general relativity $m<1$%
. However, in the brane world model, at high densities the quadratic
correction term dominates the cosmological evolution and the Hubble
parameter is proportional to the energy density, $H\sim \rho $. Therefore
values of $m$ greater than one, $m>1$, are also allowed to describe high density
bulk viscous cosmological fluids. For low density
viscous matter $s\rightarrow 1$ \cite{BeNiKh79} and
consequently on the brane $m$ could take values as high as $2$, $%
m\rightarrow 2$.

The Israel-Stewart theory is derived under the assumption that the
thermodynamic state of the fluid is close to equilibrium, which means that
the non-equilibrium bulk viscous pressure should be small when compared to
the local equilibrium pressure, that is $|\Pi |\ll p$. Consequently, the
energy density of the cosmological fluid must satisfy for all times the
condition $\rho <<\left[ \left( \gamma -1\right) /3\gamma \Pi _{0}\right]
^{1/\left( m-1\right) }$.

From\ Eq. (\ref{drho}), it follows that the time evolution of the energy
density of the matter on the brane is given by 
\begin{equation}
\rho \left( t\right) =\left[ \rho _{0}V^{\gamma \left( m-1\right) }+\Pi _{0}%
\right] ^{1/\left( 1-m\right) },  \label{d}
\end{equation}
where $\rho _{0}>0$ is a constant of integration.

In the case $m=1$ the Bianchi identity takes the form $\dot{\rho}+3\gamma
_{eff}H\rho =0$, where $\gamma _{eff}=\gamma \left( 1-\Pi _{0}\right) $.
Hence for Bianchi type I brane world filled with a viscous dissipative fluid
with bulk viscous pressure proportional to the energy density, the dynamics
is described by the same equations as for the perfect fluid case \cite
{ChHaMa01a}, with $\gamma \rightarrow \gamma _{eff}$.

Substituting ${\cal U=U}_{0}V^{-\frac{4}{3}}$ into Eqs. (\ref{dH}) and (\ref
{19}) and eliminating $\Pi $ from Eqs. (\ref{dH}) and (\ref{19}), we obtain
the basic equation for $V$ describing the dynamics and evolution of the
Bianchi type I brane world in the framework of Israel-Stewart-Hiscock causal
thermodynamics: 
\begin{equation}
6H^{2}-\frac{K^{2}}{V^{2}}=2\Lambda +2k_{4}^{2}\rho +\frac{k_{4}^{2}}{%
\lambda }\rho ^{2}+\frac{12{\cal U}_{0}}{k_{4}^{2}\lambda }V^{-\frac{4}{3}}.
\label{V}
\end{equation}

In view of the definition of the mean Hubble function $H$ and Eq. (\ref{d}),
Eq. (\ref{V}) can be integrated to give 
\begin{eqnarray}
t-t_{0} &=&\sqrt{\frac{2}{3}}\times  \nonumber \\
&&\int \left[ K^{2}+2\Lambda V^{2}+2k_{4}^{2}V^{2}\left( \rho _{0}V^{\gamma
\left( m-1\right) }+\Pi _{0}\right) ^{\frac{1}{1-m}}+\frac{k_{4}^{2}}{%
\lambda }V^{2}\left( \rho _{0}V^{\gamma \left( m-1\right) }+\Pi _{0}\right)
^{\frac{2}{1-m}}+\frac{12{\cal U}_{0}}{k_{4}^{2}\lambda }V^{\frac{2}{3}}%
\right] ^{-\frac{1}{2}}dV,  \label{t}
\end{eqnarray}
with $t_{0}$ an arbitrary constant of integration.

The scale factors $a_{i},i=1,2,3,$ are obtained from Eq. (\ref{20}), and are
given by 
\begin{equation}
a_{i}(t)=a_{i0}V^{\frac{1}{3}}e^{K_{i}\int \frac{1}{V\dot{V}}dV},i=1,2,3,
\end{equation}
where $a_{i0},i=1,2,3$ are arbitrary constants of integration.

The anisotropy parameter can be expressed in terms of $V$ and $H$ in the
form 
\begin{equation}
A(t)=\frac{1}{3}\frac{K^{2}}{V^{2}H^{2}}=2K^{2}\left[ K^{2}+2\Lambda
V^{2}+2k_{4}^{2}V^{2}\left( \rho _{0}V^{\gamma \left( m-1\right) }+\Pi
_{0}\right) ^{\frac{1}{1-m}}+\frac{k_{4}^{2}}{\lambda }V^{2}\left( \rho
_{0}V^{\gamma \left( m-1\right) }+\Pi _{0}\right) ^{\frac{2}{1-m}}+\frac{12%
{\cal U}_{0}}{k_{4}^{2}\lambda }V^{\frac{2}{3}}\right] ^{-1}.
\end{equation}

The expansion scalar $\theta $ can be represented, in a parametric form, as
a function of $V$, as 
\begin{equation}
\theta =\sqrt{\frac{3}{2}}\frac{\left[ K^{2}+2\Lambda
V^{2}+2k_{4}^{2}V^{2}\left( \rho _{0}V^{\gamma \left( m-1\right) }+\Pi
_{0}\right) ^{1/\left( 1-m\right) }+\frac{k_{4}^{2}}{\lambda }V^{2}\left(
\rho _{0}V^{\gamma \left( m-1\right) }+\Pi _{0}\right) ^{2/\left( 1-m\right)
}+\frac{12{\cal U}_{0}}{k_{4}^{2}\lambda }V^{\frac{2}{3}}\right] ^{1/2}}{V}.
\end{equation}

The shear scalar $\sigma ^{2}$ is given by 
\begin{equation}
\sigma ^{2}=\frac{K^2}{2V^{2}}.
\end{equation}

As a function of $V$ the deceleration parameter $q$ can be expressed as 
\begin{equation}
q=\frac{\left( 3\gamma -2\right) \left( \rho _{0}V^{\gamma \left( m-1\right)
}+\Pi _{0}\right) ^{1/\left( 1-m\right) }-3\gamma \Pi _{0}\rho ^{m}+2\frac{%
K^{2}}{V^{2}}}{2\left[ \left( \rho _{0}V^{\gamma \left( m-1\right) }+\Pi
_{0}\right) ^{1/\left( 1-m\right) }+\frac{K^{2}}{2V^{2}}\right] }.
\end{equation}

Integrating Eq. (\ref{bulk}), we obtain the time dependence of the
temperature of the matter on the brane in the form 
\begin{equation}
T\left( t\right) =T_{0}\frac{\tau \Pi ^{2}}{\xi }\exp \left[ 2\int \left( 
\frac{1}{\tau }+\frac{3\xi H}{\tau \Pi }+\frac{3H}{2}\right) dt\right] ,
\label{bulk1}
\end{equation}
with $T_{0}$ an arbitrary constant of integration.

Following Belinskii, Nikomarov and Khalatnikov \cite{BeNiKh79}, we suppose
that, in order to guarantee that the propagation velocity of bulk viscous
perturbations, i.e., the non-adiabatic contribution to the speed of sound $%
c_b$ does not exceed the speed of light, the relation $\tau =\xi
/c_{b}^{2}(\rho +p)=\xi /\gamma c_{b}^{2}\rho $ holds in the dissipative
fluid. With this choice for the relaxation time $\tau $ the causal evolution
Eq. (\ref{bulk1}) gives the following evolution law for the temperature: 
\begin{equation}
T(t)=T_{0}\gamma \frac{\Pi _{0}^{2}}{c_{b}^{2}}\rho ^{2m-1}\exp \left[ 2\int
\left( \gamma c_{b}^{2}\frac{\rho }{\xi }-3\frac{\gamma c_{b}^{2}}{\Pi _{0}}%
\rho ^{1-m}H+\frac{3H}{2}\right) dt\right] .  \label{20a}
\end{equation}

An analysis of the relativistic kinetic equation for some simple cases given
by Murphy \cite{Mu73}, Belinskii and Khalatnikov \cite{BeKh75} and
Belinskii, Nikomarov and Khalatnikov \cite{BeNiKh79} has shown that in the
asymptotic regions of small and large values of the energy density, the
viscosity coefficients can be approximated by power functions of the energy
density with definite requirements on the exponents of these functions. For
small values of the energy density it is reasonable to consider large
exponents, equal in the extreme case to one. For large $\rho $ the power of
the bulk viscosity coefficient should be considered smaller (or equal) to $%
1/2$. Hence, we shall assume that the bulk viscosity coefficient obeys the
simple phenomenological law 
\begin{equation}
\xi (t)=\alpha \rho ^{s},  \label{21a}
\end{equation}
with $\alpha \ge 0$ and $0\le s\le 1/2$ constants.

Therefore we obtain for the temperature of the dissipative viscous
cosmological fluid on the brane the following representation: 
\begin{eqnarray}
T &=&T_{0}\gamma \frac{\Pi _{0}^{2}}{c_{b}^{2}}\rho ^{2m-1}\exp \left[ 2\int
\left( \frac{\gamma c_{b}^{2}}{\alpha }\rho ^{1-s}-3\frac{\gamma c_{b}^{2}}{%
\Pi _{0}}\rho ^{1-m}H+\frac{3H}{2}\right) dt\right] =T_{0}\gamma \frac{\Pi
_{0}^{2}}{c_{b}^{2}}\rho ^{2m-1}\times  \nonumber \\
&&\exp \left[ \sqrt{\frac{8}{3}}\int \frac{\left( \frac{\gamma c_{b}^{2}}{%
\alpha }\rho ^{1-s}-3\frac{\gamma c_{b}^{2}}{\Pi _{0}}\rho ^{1-m}H+\frac{3H}{%
2}\right) }{\left[ K^{2}+2\Lambda V^{2}+2k_{4}^{2}V^{2}\left( \rho
_{0}V^{\gamma \left( m-1\right) }+\Pi _{0}\right) ^{\frac{1}{1-m}}+\frac{%
k_{4}^{2}}{\lambda }V^{2}\left( \rho _{0}V^{\gamma \left( m-1\right) }+\Pi
_{0}\right) ^{\frac{2}{1-m}}+\frac{12{\cal U}_{0}}{k_{4}^{2}\lambda }V^{%
\frac{2}{3}}\right] ^{\frac{1}{2}}}dV\right].
\end{eqnarray}

The growth of the total comoving entropy $\Sigma (t)$ over a proper time
interval $(t_{0},t)$ is given by 
\begin{equation}
\Sigma (t)-\Sigma (t_{0})=-\frac{3}{k_{B}}\int_{t_{0}}^{t}\frac{\Pi VH}{T}%
dt=-\frac{1}{k_{B}}\int_{V_{0}}^{V}\frac{\Pi }{T}dV,  \label{entrop}
\end{equation}
where $k_{B}$ is the Boltzmann constant.

\section{Discussions and final remarks}

In the present paper we have considered the evolution of a causal viscous
dissipative cosmological fluid in the brane world scenario. As only source
of dissipation we have considered the bulk viscosity of the matter on the
brane. In this case the standard Einstein equations are modified, due to the
presence of the terms from extra dimensions, quadratic in the energy density.

By assuming a power-law dependence of the bulk viscous pressure on the
energy density of the cosmological fluid, the general solution of the field
equations for an anisotropic Bianchi type I geometry can be obtained in an
exact parametric form. By assuming for the bulk viscosity coefficient and
for the relaxation time the usual equations of state, the temperature
evolution of the cosmological fluid can also be obtained in a closed form.

The cosmological evolution of the bulk viscous brane Universe is
expansionary, with all the scale factors monotonically increasing functions
of time. For $t\rightarrow \infty $ we have $a_{i}=a=a_{0}V^{1/3}$, $i=1,2,3$%
. Hence the Universe ends in an isotropic state. The expansion parameter $%
\theta $ is a decreasing function of time, with a singular behavior at $t=0$.

The time evolution of the energy density of the cosmological fluid
essentially depends on the value of the exponent $m$ in the bulk viscous
pressure-energy density relation. The variation with respect the
cosmological time of $\rho $ is represented, for $m>1$, in Fig. 1.

\begin{figure}[h]
\epsfxsize=10cm
\centerline{\epsffile{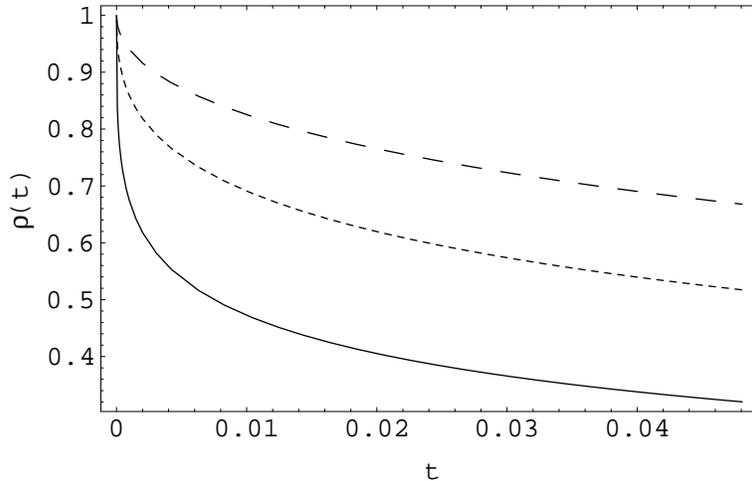}}
\caption{Time evolution of the energy density $\rho (t)$ of the
Zeldovich fluid ($\gamma =2$) filled Bianchi type I brane Universe,
for different values of the parameter $m$: $m=1.25$ (solid curve), $m=1.45$
(dotted curve) and $m=1.6$ (dashed curve). We have normalized the constants
so that $K^{2}=2$, $\Lambda =0.5$, $\rho _{0}=\Pi _{0}$, $%
2k_{4}^{2}rho _{0}^{1/(1-m)}=1$, $k_{4}^{2}\rho _{0}^{2/(1-m)}=%
\lambda $ and $12{\cal U}_{0}=k_{4}^{2}\lambda $.}
\label{FIG1}
\end{figure}

For this range of values of $m$ the Universe starts its evolution with
finite values of the energy density, $\rho (0)=\Pi _{0}^{1/(1-m)}\neq 0$.
But the scale factors are singular at the initial time, $a_i(t_0)=0$, $%
i=1,2,3$. For $m<1$ the anisotropic brane Universe starts from a singular
state, with infinite energy density and zero scale factors. For all values
of $m$ the energy density is a monotonically decreasing function of time.

The evolution of the deceleration parameter $q$ is represented in Fig. 2.

\begin{figure}[h]
\epsfxsize=10cm
\centerline{\epsffile{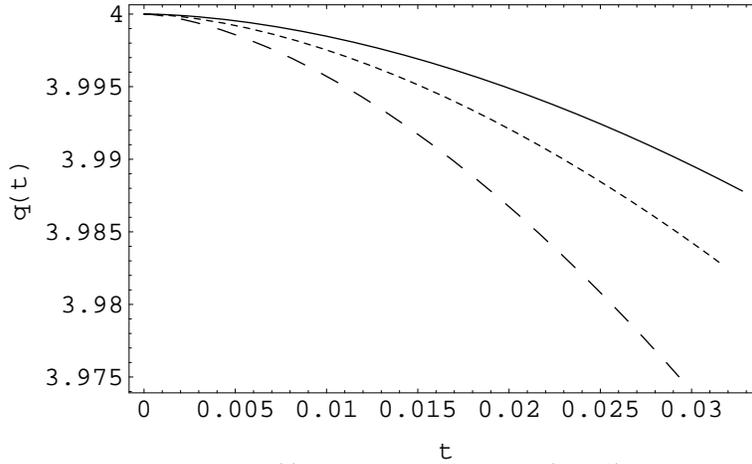}}
\caption{Dynamics of the deceleration parameter $q(t)$ of the Zeldovich
fluid ($\gamma =2$) filled Bianchi type I brane Universe, for
different values of the parameter $m$: $m=1.3$ (solid curve), $m=1.4$
(dotted curve) and $m=1.5$ (dashed curve). We have normalized the constant
parameters so that $K^{2}=2$, $\Lambda =0.5$, $\rho _{0}=\Pi _{0}$, $%
2k_{4}^{2}\rho _{0}^{1/(1-m)}=1$, $k_{4}^{2}\rho
_{0}^{2/(1-m)}=\lambda $ and $12{\cal U}_{0}=k_{4}^{2}\lambda
$.}
\label{FIG2}
\end{figure}

Due to the smallness of the bulk viscous pressure, satisfying the condition $%
\left| \Pi \right| <<p$, the cosmological evolution is non-inflationary,
with $q>0$ for all values of $m$ and for all $t$. In order to have
inflationary evolution it is necessary that the negative bulk viscous
pressure dominates the thermodynamic pressure.

The dynamics of the mean anisotropy parameter $A$ strongly depends on the
value of $m$. The time variation of $A$ for $m<1$ is represented in Fig. 3.
In this case the Bianchi type I Universe starts its evolution from an
isotropic (but singular) state, with $A(0)=0$. For small times the mean
anisotropy is increasing, reaching a maximum at a finite time $t=t_{c}$. For
times $t>t_{c}$ the anisotropy is a monotonically decreasing function of
time, and thus in the large time limit the anisotropic brane world Universe
will end in an homogeneous and isotropic state. As has already been pointed
out in \cite{ChHaMa01a}, this behavior is specific for brane world
cosmological models and cannot be found in conventional general relativistic
scenarios. For $m>1$, the brane Universe starts from a state of maximum
anisotropy, also reaching in the long time limit an isotropic phase. Due to
the expansion of the Universe, the shear scalar $\sigma ^{2}$ vanishes in
the large time limit, $\sigma ^{2}\rightarrow 0$ for $V\rightarrow \infty $.

\begin{figure}[h]
\epsfxsize=10cm
\centerline{\epsffile{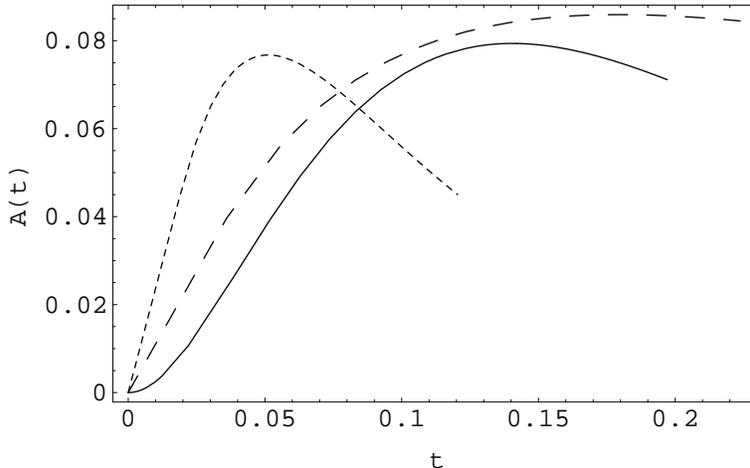}}
\caption{Evolution of the mean anisotropy parameter $A(t)$ of the Zeldovich
fluid ($\gamma =2$) filled Bianchi type I brane Universe, for
different values of the parameter $m$: $m=0.7$ (solid curve), $m=0.8$
(dotted curve) and $m=0.9$ (dashed curve). We have normalized the constants
so that $K^{2}=2$, $\Lambda =0.5$, $\rho _{0}=\Pi _{0}$, $2k_{4}^{2}%
\rho _{0}^{1/(1-m)}=1$, $k_{4}^{2}\rho _{0}^{2/(1-m)}=%
\lambda $ and $12{\cal U}_{0}=k_{4}^{2}\lambda $.}
\label{FIG3}
\end{figure}

According to the formal definition of the isotropization given by
Collins and Hawking \cite{CoHa73} a cosmological model approaches isotropy
if the following four conditions hold as $t\rightarrow \infty $: i) the
Universe is expanding indefinetely and $H>0$ ii) $T^{00}>0$ and $%
T^{0i}/T^{00}\rightarrow 0$. $T^{0i}/T^{00}$ represents an average velocity
of the matter relative to the surfaces of homogeneity. If this does not tend
to zero, the Universe would not appear homogeneous or isotropic iii) the
anisotropy in the locally measured Hubble constant $\sigma /H$ tends to
zero, $\sigma /H\rightarrow 0$ and iv) the distortion part of the metric
tends to a constant. All these conditions are satisfied in the present
model, since in the large time limit the bulk viscous brane Universe is
expanding and the energy density of the matter is positive for all times.
Moreover, as one can immediately see from Eq. (\ref{10}) the quantity
$\sigma /H$ is proportional to the square root of the anisotropy parameter,
$\sigma /H\sim \sqrt{A}$ and in the large time for all values of the
parameter $m$ it tends to zero, $\sigma /H\rightarrow 0$. As for the
condition iv), it is also satisfied since in the metric we have considered
only the volume part. Spatially homogeneous models can be divided in three
classes: those which have less than the escape velocity (i.e., those whose
rate of expansion is insufficient to prevent them from recollapsing), those which
have just the escape velocity and those which have more than the escape
velocity \cite{CoHa73}. Models of the third class do not tend, generally, to isotropy.
In fact the only types which can tend toward isotropy at
arbitrarily large times are types $I$, $V$, $VII_{0}$ and $VII_{h}$. For
type $VII_{h}$ there is no nonzero measure set of these models which tends
to isotropy \cite{CoHa73}. In this sense the Bianchi type I model we have studied
in the present paper is a very special case. The Bianchi types that drive flat
and open Universes away from isotropy in the Collins-Hawking sense are those of
type $VII$. 

As a consequence of its energy density dependence, the bulk viscous pressure
is a decreasing function of time for all values of the parameters and for
large times it tends to zero. The time evolution of $\Pi $ essentially
depends on the constant $m$. For $m>>1$, the time scale in which $\Pi
\approx 0$ is very short.

The variation of the temperature, presented in Fig. 4, shows that at the
initial times of the cosmological evolution the temperature is an increasing
function of time, reaching a maximum value at a finite time $t_{c}$. For
times $t>t_{c}$, the temperature is a decreasing function of time.

\begin{figure}[h]
\epsfxsize=10cm
\centerline{\epsffile{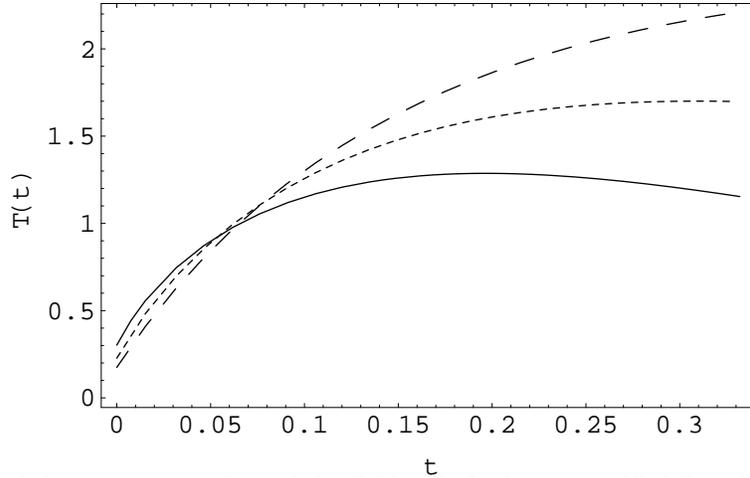}}
\caption{Time evolution of the temperature $T(t)$ of the Zeldovich fluid ($%
\gamma =2$) filled Bianchi type I brane Universe, for $s=1/4$ and
for different values of the parameter $m$: $m=0.45$ (solid curve), $m=0.65$
(dotted curve) and $m=0.9$ (dashed curve). We have normalized the constants
so that $K^{2}=2$, $\Lambda =0.5$, $\rho _{0}=\Pi _{0}$, $2k_{4}^{2}%
\rho _{0}^{1/(1-m)}=1$, $k_{4}^{2}\rho _{0}^{2/(1-m)}=%
\lambda $, $12{\cal U}_{0}=k_{4}^{2}\lambda $, $%
\gamma T_{0}\Pi _{0}^{2}=c_{b}^{2}$, $\gamma c_{b}^{2}=\alpha
$ and $3c_{b}^{2}=\Pi _{0}$ .}
\label{FIG4}
\end{figure}

The comoving entropy, represented in Fig. 5, is an increasing function of
time in the early stages of cosmological evolution, but it tends to a
constant value in the large time limit.

\begin{figure}[h]
\epsfxsize=10cm
\centerline{\epsffile{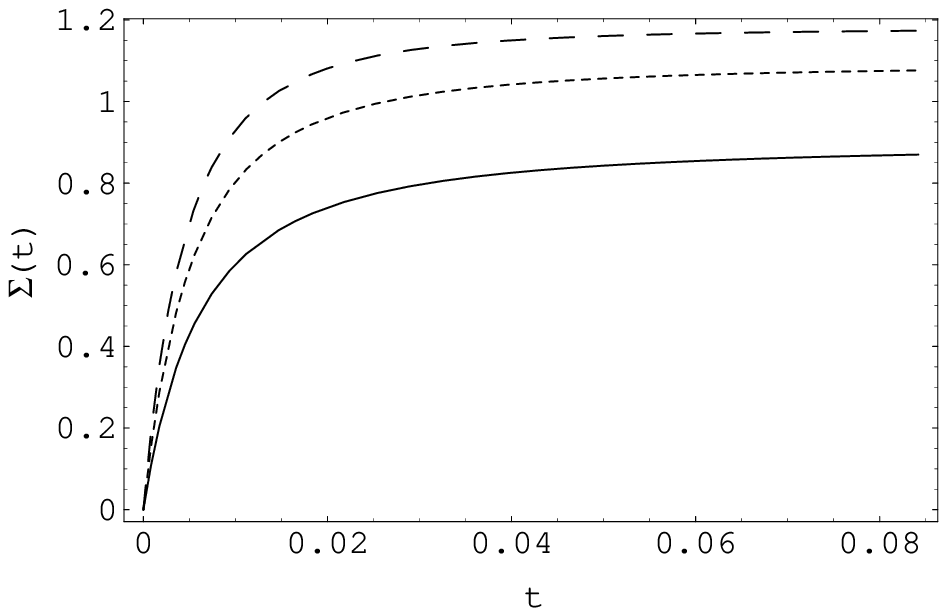}}
\caption{Variation, as a function of time, of the comoving entropy $\Sigma
(t)$ of the Zeldovich fluid ($\gamma =2$) filled Bianchi type I
brane Universe, for $s=1/4$ and for different values of the parameter $m$: $%
m=0.45$ (solid curve), $m=0.65$ (dotted curve) and $m=0.9$ (dashed curve).
We have normalized the constants so that $K^{2}=2$, $\Lambda =0.5$, $%
\rho _{0}=\Pi _{0}$, $2k_{4}^{2}\rho _{0}^{1/(1-m)}=1$, $k_{4}^{2}%
\rho _{0}^{2/(1-m)}=\lambda $, $12{\cal U}_{0}=k_{4}^{2}%
\lambda $, $\gamma T_{0}\Pi _{0}^{2}=c_{b}^{2}$, $%
\gamma c_{b}^{2}=\alpha $ and $3c_{b}^{2}=\Pi _{0}$ .}
\label{FIG5}
\end{figure}

The transition from the anisotropic to the isotropic state of the brane
world is associated to a large increase in the entropy. The increase of the
comoving entropy of the cosmological fluid is independent of the value of
the parameter $m$ and is a general feature of the model. In the large time
limit $\Pi \rightarrow 0$ and, as one can see from Eq. (\ref{entrop}), in
this limit $\Sigma \rightarrow const.$

The equation of state of the Zeldovich fluid with $\gamma =2$ is equivalent
to the equation of state of a massless scalar field $\phi $, with $\rho =p=%
\dot{\phi}^{2}/2$ and a dissipative coupling \cite{Ba87}, with $\phi $
satisfying the equation
\begin{equation}
\ddot{\phi}+3H\dot{\phi}=3\left( 2^{1-m}\right) \Pi _{0}\dot{\phi}^{2m-1}H.
\end{equation}

The evolution of the scalar field can be obtained again in parametric form.
Classically, bulk viscous stresses arise because the expansion of the
Universe is continually trying to pull the fluid out of thermal equilibrium,
and the cosmological fluid is trying to relax back. Hence the viscosity of a classical fluid
arises from the differential motion of fluid elements. But in the collisionless
periods after the Planck era and during inflation there are no particle
interactions responsible for the viscosity. However, the phenomenological
description of quantum particle production processes can be consistently
modeled by means of bulk viscosities \cite{MaHa99a}, \cite{Hu82}. In
particular, the full causal thermodynamics can be reformulated as a theory
describing particle production \cite{MaHa99a}. In this formulation the
causal bulk viscous pressure $\Pi $ in the energy conservation equation (\ref
{drho}) acts as a creation pressure $p_{c}$, $p_{c}=\Pi $, while the
particle production rate $\Gamma $ is proportional to the bulk viscous
pressure divided by the locally measured Hubble parameter. The particle
balance equation, describing the variation of the particle number due to
the combined effects of the expansion of the Universe and particle production, 
takes the form
\begin{equation}
\dot{n}+3Hn=\Gamma n,
\end{equation}
where $\Gamma =-\Pi /\gamma H$ \cite{MaHa99a}.

As an example of particle creation processes on the brane we consider
particle production from a scalar field obeying an equation of state of the
form $p_{\phi }=\left( \gamma _{\phi }-1\right) \rho _{\phi }$, $0\leq
\gamma _{\phi }\leq 1$, where $\rho _{\phi }=\dot{\phi}^{2}/2+U\left( \phi
\right) $ and $p_{\phi }=\dot{\phi}^{2}/2-U\left( \phi \right) $, with $%
U\left( \phi \right) $ the self-interaction potential. The deviation from
the conformal invariance of the field or the polarization (trace anomaly) of
quantized fields interacting with dynamic spacetimes will lead to particle production \cite
{Hu82}. The rate of particle creation in a particular mode depends on its
frequency in relation to the expansion rate of the Universe: at any given
time $t$ in the quantum regime, new particles with frequency $\omega \leq
t^{-1}$ are created from the quantum vacuum, while particles produced
earlier interact and have an expansionary dynamics. Whether the newly
created particles can reach effective thermal equilibrium depends on the
rate of their interaction relative to the rate of expansion.

To obtain a phenomenological classical description of the essentially
quantum creation process, we assume again that the negative creation
pressure is proportional to the energy density of the scalar field, $%
p_{c}=-p_{c0}\gamma _{\phi }\rho _{\phi }^{m}$, where, for simplicity, we
suppose that all $p_{c0}$, $\gamma _{\phi }$ and $m$ are constants. Then the
time variation of the scalar field is given by 
\begin{equation}
\rho _{\phi }(t)=\left[ \rho _{0\phi }V^{\gamma _{\phi }\left( m-1\right)
}+p_{c0}\right] ^{1/\left( 1-m\right) },
\end{equation}
where $\rho _{0\phi }\geq 0$ is a constant of integration. The rate at which
particles are created on the brane is 
\begin{equation}
\Gamma (t)=p_{c0}H^{-1}\left[ \rho _{0\phi }V^{\gamma _{\phi }\left(
m-1\right) }+p_{c0}\right] ^{m/\left( 1-m\right) },
\end{equation}
while the particle number varies as
\begin{equation}
n(t)=\frac{n_{0}}{V}\exp \left( \int \Gamma (t)dt\right) ,
\end{equation}
with $n_{0}\geq 0$ a constant of integration. At high temperature $n=g\zeta
\left( 3\right) T^{3}/\pi ^{2}$, where
$\zeta \left( s\right) $ is the Riemann function, $\zeta \left(
s\right) =\sum_{k=1}^{\infty }k^{-s}$, with $\zeta \left( 3\right) =1.202$ and $g$
is the total spin state of the particles. Therefore the temperature of the
very early Universe is given by
\begin{equation}
T(t)=\left( \frac{n_{0}\pi ^{2}}{g\zeta \left( 3\right) }\right) ^{1/3}\frac{%
\exp \left( \frac{1}{3}\int \Gamma (t)dt\right) }{V^{1/3}}.
\end{equation}

In the particular case $\gamma _{\phi }=0$, corresponding to a dissipative
scalar field obeying an equation of state of the form $\rho _{\phi }+p_{\phi
}=1`0$, the time variation of the energy density of the field is $\rho _{\phi
}\sim t^{1/\left( 1-m\right) }$, while the creation pressure varies as $%
p_{c}\sim t^{m/(1-m)}$. In order to have a decaying field it is necessary
that $m>1$. Even in the presence of particle creation processes, the initial
expansion of the brane Universe is non-inflationary. But in the large time
limit, for large values of $V$ and for a decaying creation pressure with $%
p_{c}\rightarrow 0$, the deceleration parameter becomes $q=\left[ \left(
3\gamma _{\phi }-2\right) \right] /2$. For $\gamma _{\phi }<2/3$, the brane
Universe ends in an inflationary epoch.  

Since the dynamics on the brane is different from the standard general
relativistic one, the particle creation processes are also influenced by the
effects of the extra dimensions. Because the rate of expansion is higher,
the particle creation processes are also accelerated in the models with
large extra dimensions. 

In the present paper we have pointed out an other important difference
between brane world cosmology and standard general relativity predictions
about the very early Universe, namely the behavior of the temperature of a
realistic cosmological fluid. In brane world cosmological models, the high
density Universe starts from a low temperature state. In the initial stages
of expansionary evolution the increase of the degree of anisotropy is
associated to an increase of the temperature of the Universe, the
cosmological fluid experiencing a heating process. An other interesting
feature of the model is the presence of a non-singular energy density,
associated to a singularity of the scale factors. The present approach can
also lead to a better understanding of the entropy generation mechanisms in
the very early Universe, where viscous type physical processes have probably
played an important role even when the Universe was about $1000s$ old \cite
{Ma95}.

Hence dissipative viscous brane world cosmological model with an anisotropic
geometry could maybe describe a well-determined period of the very early
evolution of our Universe.

\section*{Acknowledgments}

The authors would like to thank to the two anonymous referees, whose comments and suggestions helped  to significantly
improve the manuscript.


\end{document}